\documentclass{elsart}%
\usepackage{amsmath}
\usepackage{graphicx}%
\usepackage{amsfonts}%
\usepackage{amssymb}

\begin{document}
\begin{frontmatter}
\title{On size and growth of business firms}%

\author[ATA]{G. De Fabritiis}
\author[DES]{F. Pammolli \corauthref{cor1}%
}
\author[DES]{M. Riccaboni}%

\address[ATA]{ ATA - Advanced Technology Assessment,\\
 Via Stradivari 19, 50127 Florence, Italy}%

\address[DES]{ DSA, University of Florence,\\
 Via Montebello 7, 50123 Florence, Italy }%

 \corauth[cor1]{Corresponding author. E-mail address: pammolli@cln.it\\
(F. Pammolli); Tel. +39 0577 232195; Fax: +39 0577 232188.}%
\begin{abstract}%

We study size and growth distributions of products and business
firms in the context of a given industry. Firm size growth is
analyzed in terms of two basic mechanisms, i.e. the increase of
the number of new elementary business units  and their size
growth. We find a power-law relationship between size and the
variance of growth rates for both firms and products, with an
exponent between -0.17 and -0.15, with a remarkable stability upon
aggregation. We then introduce a simple and general model of
proportional growth for both the number of firm independent
constituent units and their size, which conveys a good
representation of the empirical evidences. This general and
plausible generative process can account for the observed scaling
in a wide variety of economic and industrial systems. Our findings
contribute to shed light on the mechanisms that sustain economic
growth in terms of the relationships between the size of economic
entities and the number and size distribution of their elementary
components.
\end{abstract}%


\end{frontmatter}

\section{Introduction}

This work is rooted in the ``old'' stochastic tradition of the
analysis of economic and industrial growth \cite{Ijiri_Simon77}.
We elaborate on some recent contributions
\cite{stanley96,sutton02,gabaix02}, focusing on the shape and
width of growth rates distributions. Size and growth distributions
for firms and products are analyzed in the context of the
worldwide pharmaceutical industry, over a period of ten years. We
consider the entire population of firms and products, as well as
entry and exit of firms and products. In accordance with
\cite{stanley96} and \cite{sutton02}, we find the distribution of
growth rates to be non-Gaussian with heavy tails. Moreover, for
both products and firms, the width of the growth distribution
scales as a power law of size, with a scaling exponent $\beta$
between -0.17 and -0.15, which is remarkably stable upon
aggregation. We introduce a general framework to account for the
observed regularities, drawing some general implications on the
mechanisms which sustain economic and industrial growth. We show
that \cite{Ijiri_Simon77,Ijiri_Simon75} can be extended to account
for the shape of size and growth distributions, as well as for
scaling relationships at different levels of aggregation. Products
are considered as business opportunities which are captured and
lost by firms, and then grow in size. Both the capture and loss of
business opportunities are modelled as an instantiation of the Law
of Proportionate Effect applied to elementary business units.
Then, each elementary unit is assumed to grow in size according to
a process of proportional growth based on  random multiplicative
dynamics, with shocks independently and randomly drawn from a
lognormal distribution. This simple and general framework accounts
for the most salient features of size and growth distributions at
different levels of aggregation.

\section{Empirical findings}

Data used in this work are drawn from the Pharmaceutical Industry Database
(PHID) at CERM/EPRIS. PHID records quarterly sales figures of 48,819
pharmaceutical products commercialized by 3,919 companies in the European
Union and North America from September 1991 to June 2001 (values are in
Sterling at constant 2001 exchange rates). Information is available on entry
and exit of firms and products over time, and the entire size distribution is covered.

$S_{m,i,j}(t)$ denotes sale figures at time $t$ for the market ($m$), at the
level of each firm ($i$), and at the product level ($j$), respectively:%
\[
S_{m}(t)=%
%TCIMACRO{\tsum \limits_{i=1}^{M}}%
%BeginExpansion
{\textstyle\sum\limits_{i=1}^{M}}
%EndExpansion
S_{i}(t)=\overset{M}{\underset{i=1}{%
%TCIMACRO{\tsum }%
%BeginExpansion
{\textstyle\sum}
%EndExpansion
}}\underset{j=1}{\overset{N_{i}}{%
%TCIMACRO{\tsum }%
%BeginExpansion
{\textstyle\sum}
%EndExpansion
}}S_{ij}(t)
\]
where $M$ is the total number of firms active at time $t$ and $N_{i}$ is the
number of products of the $i$-th firm. Throughout the paper we focus on firm
internal growth. We study the logarithm of growth rates:%
\[
g_{m,i,j}(t)=\log(G_{m,i,j}(t))=\log\left(  \frac{S_{m,i,j}(t)}{S_{m,i,j}%
(t-1)}\right)  =s_{m,i,j}(t)-s_{m,i,j}(t).
\]

Market size has more than doubled from 61.500 to 159.000 \pounds M
from 1991 to 2001 (+154\%). The growth of the market was sustained
by the entry of new products. The number of products has increased
linearly from about 25,000 to more than 35,000, while the number
of firms has remained almost constant around 2,000. Both the mean
and the standard deviation of the number of products by firm have
increased linearly in time, from 12.5 to about 17 and from 44 to
about 54, respectively. The rapid market expansion of the market
notwithstanding, the first two moments of the logarithm of size
and growth distributions have been almost stationary, apart from a
marked seasonality (see Fig. 1.b-c). Both product and firm growth
distributions are non-Gaussian and leptokurtotic (Fig. 2a).Size
distributions for both products and firms are consistent with a
log-normal fit (Fig. 2b) apart from a pronounced Pareto upper tail
$1/S$.

\begin{figure}[!t]
\begin{center}
\includegraphics[height=5cm,width=6.5cm]{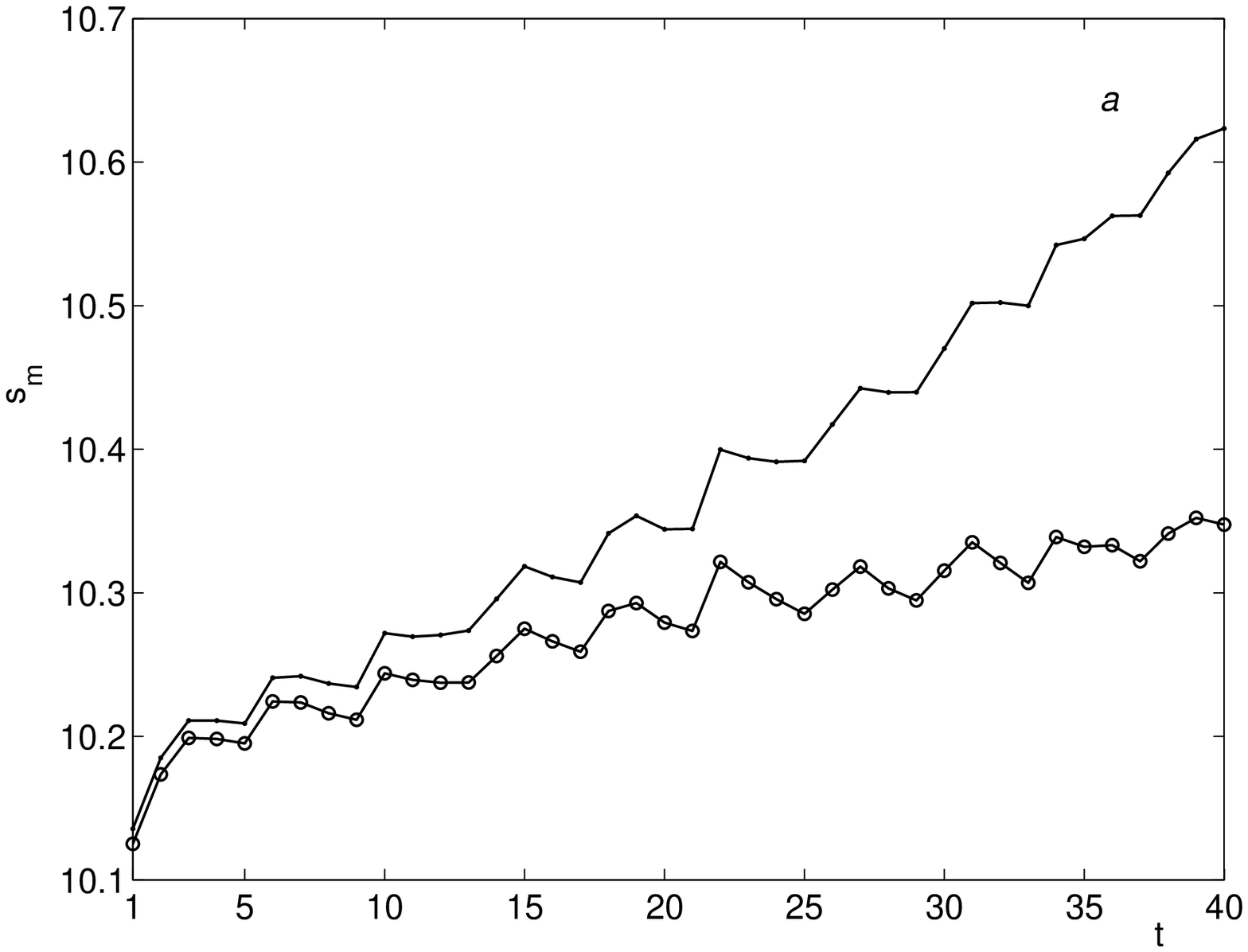}
\includegraphics[height=5cm,width=6.5cm]{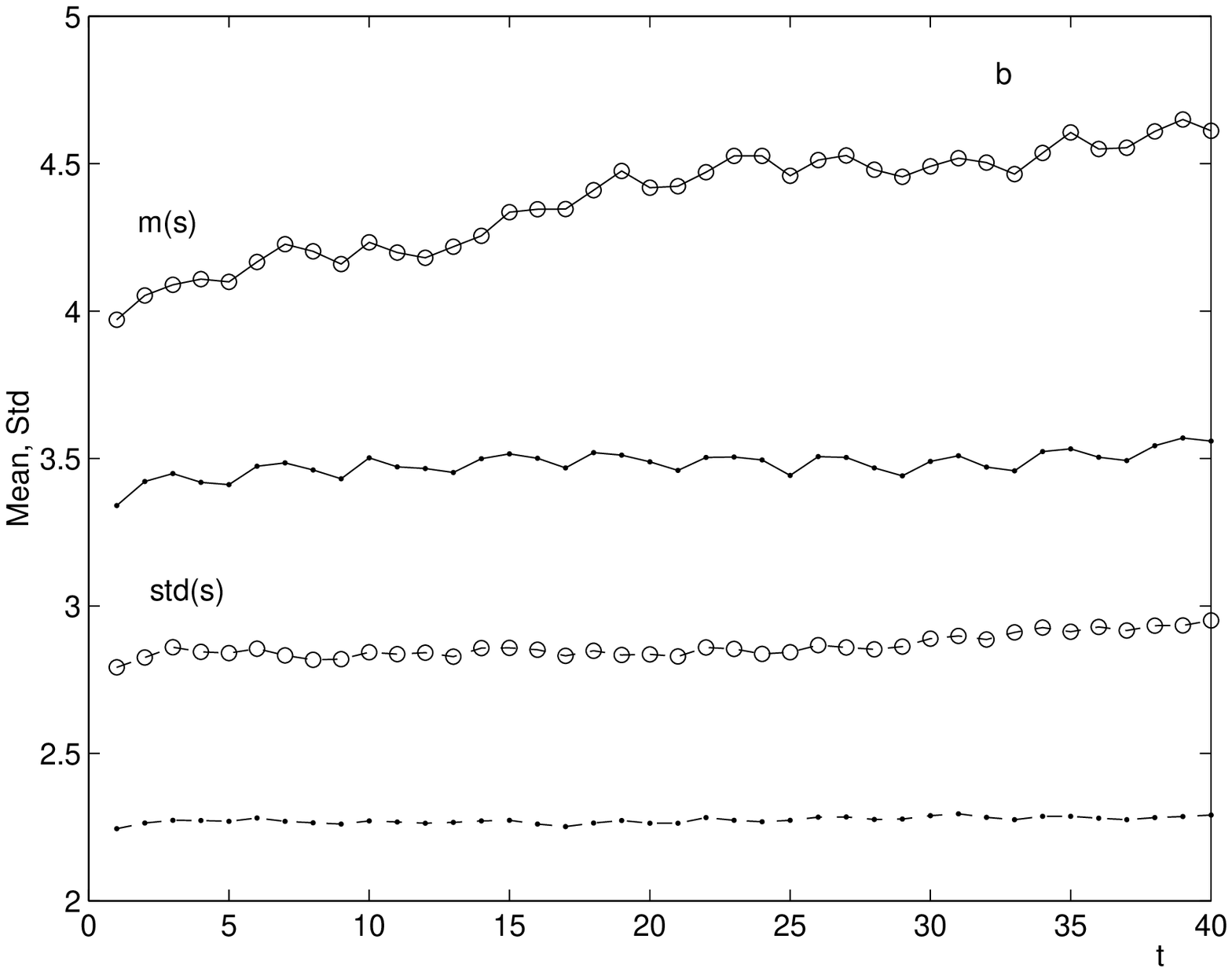}  \newline
\includegraphics[height=5cm,width=6.5cm]{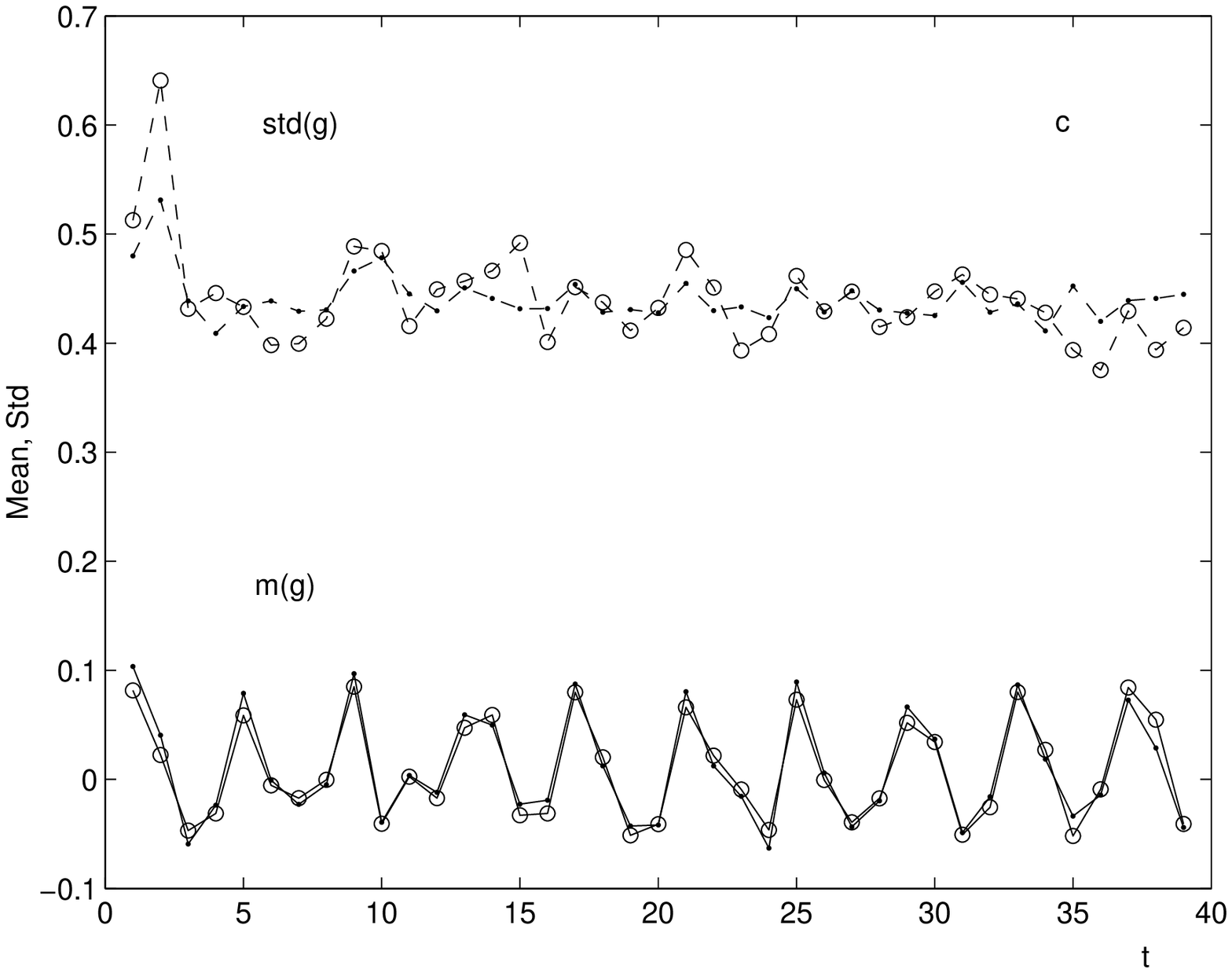}
\includegraphics[height=5cm,width=6.5cm]{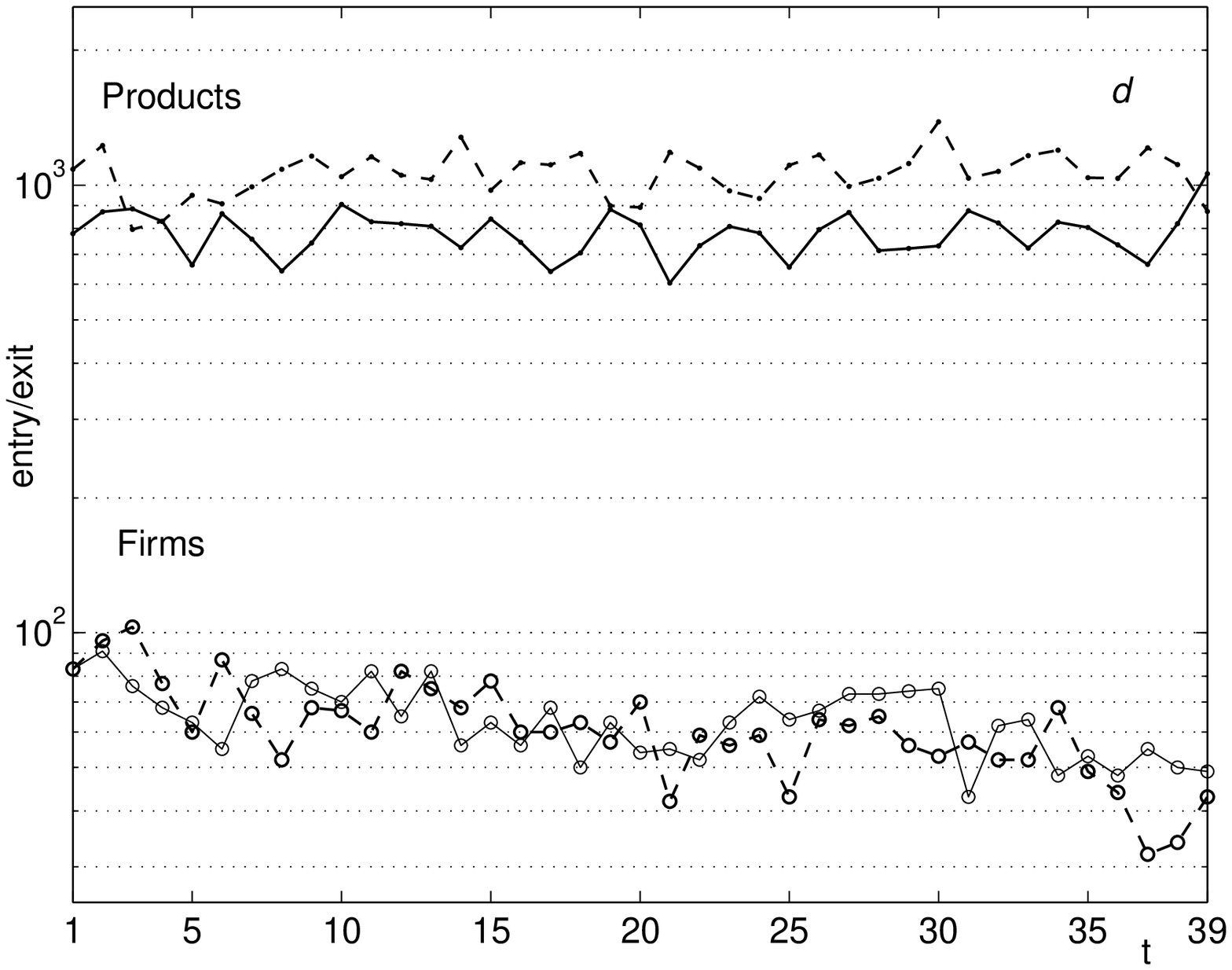}
\end{center}
\caption{Descriptive statistics. Fig. 1.a - Overall market size (dots) and
market size without entry of new products (circles). Fig. 1.b - Product (dots)
and firm (circles) size; mean and standard deviation. Fig. 1.c - Product
(dots) and firm (circles) growth; mean and standard deviation. Fig. 1.d -
Entry (full line) and exit (dashed line) patterns of products (dots) and firms
(circles). All values are in logarithmic scale.}%
\end{figure}

\begin{figure}[!t]
\begin{center}
\includegraphics[height=5cm,width=6.5cm]{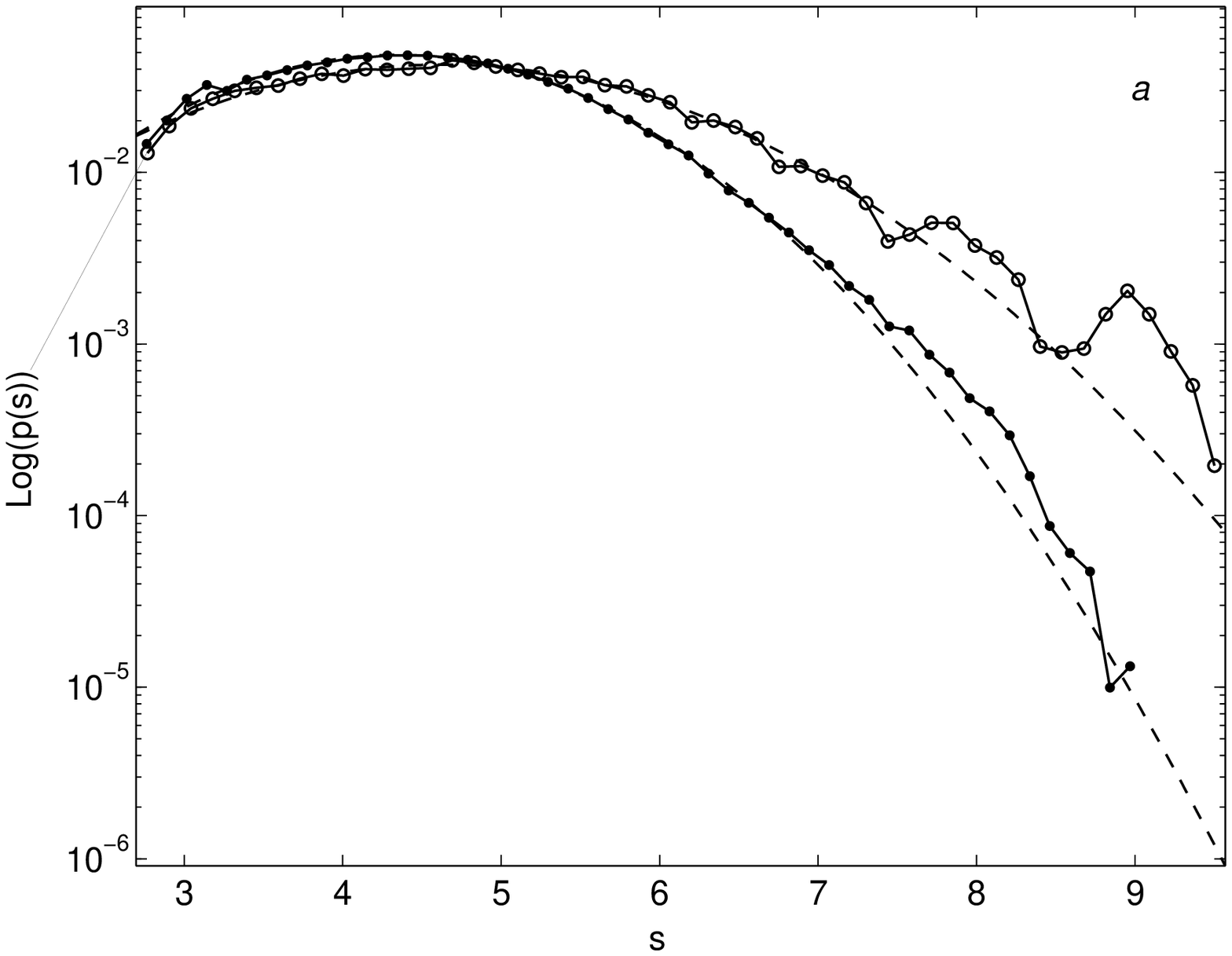}
\includegraphics[height=5cm,width=6.5cm]{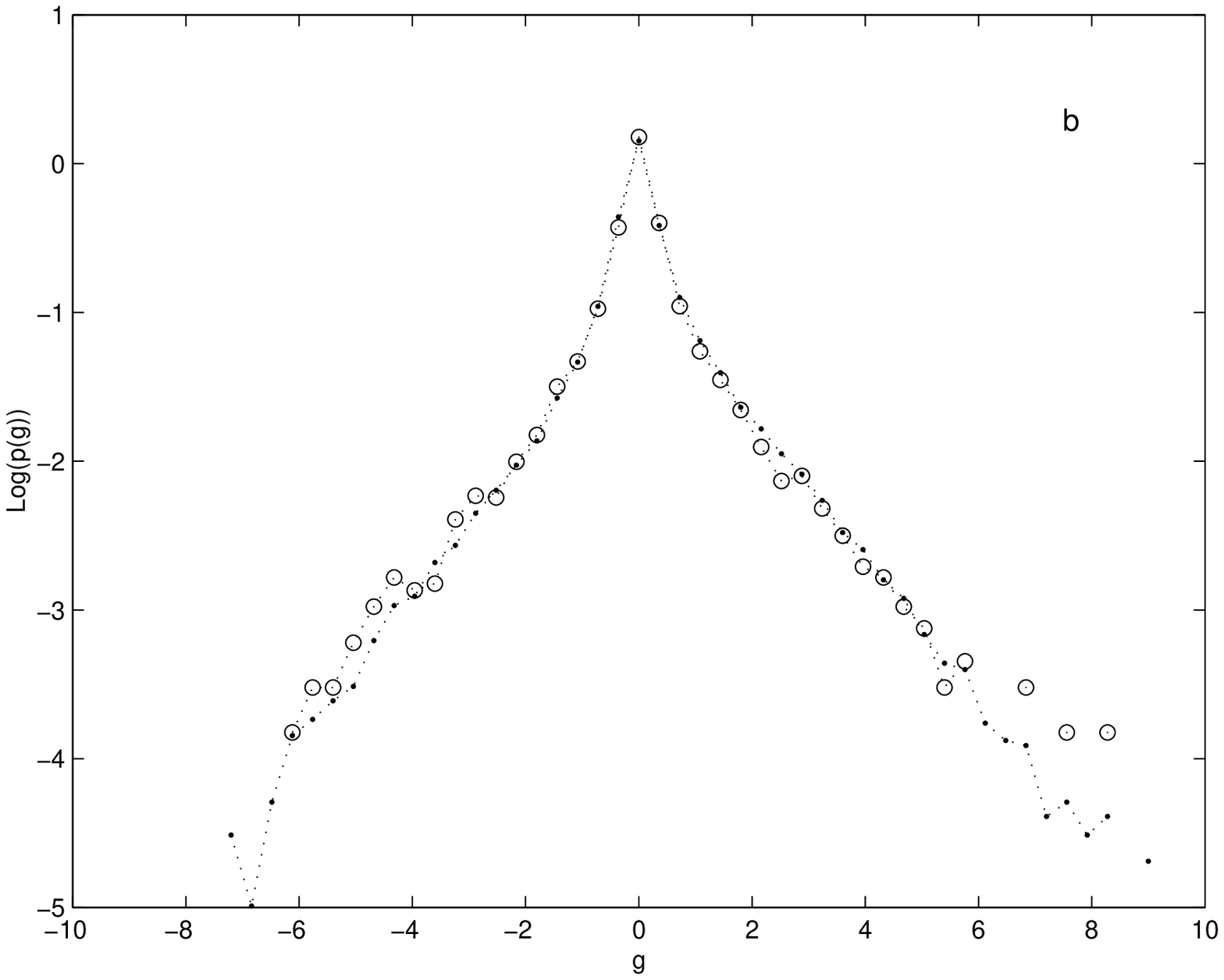}  \newline
\includegraphics[height=5cm,width=6.5cm]{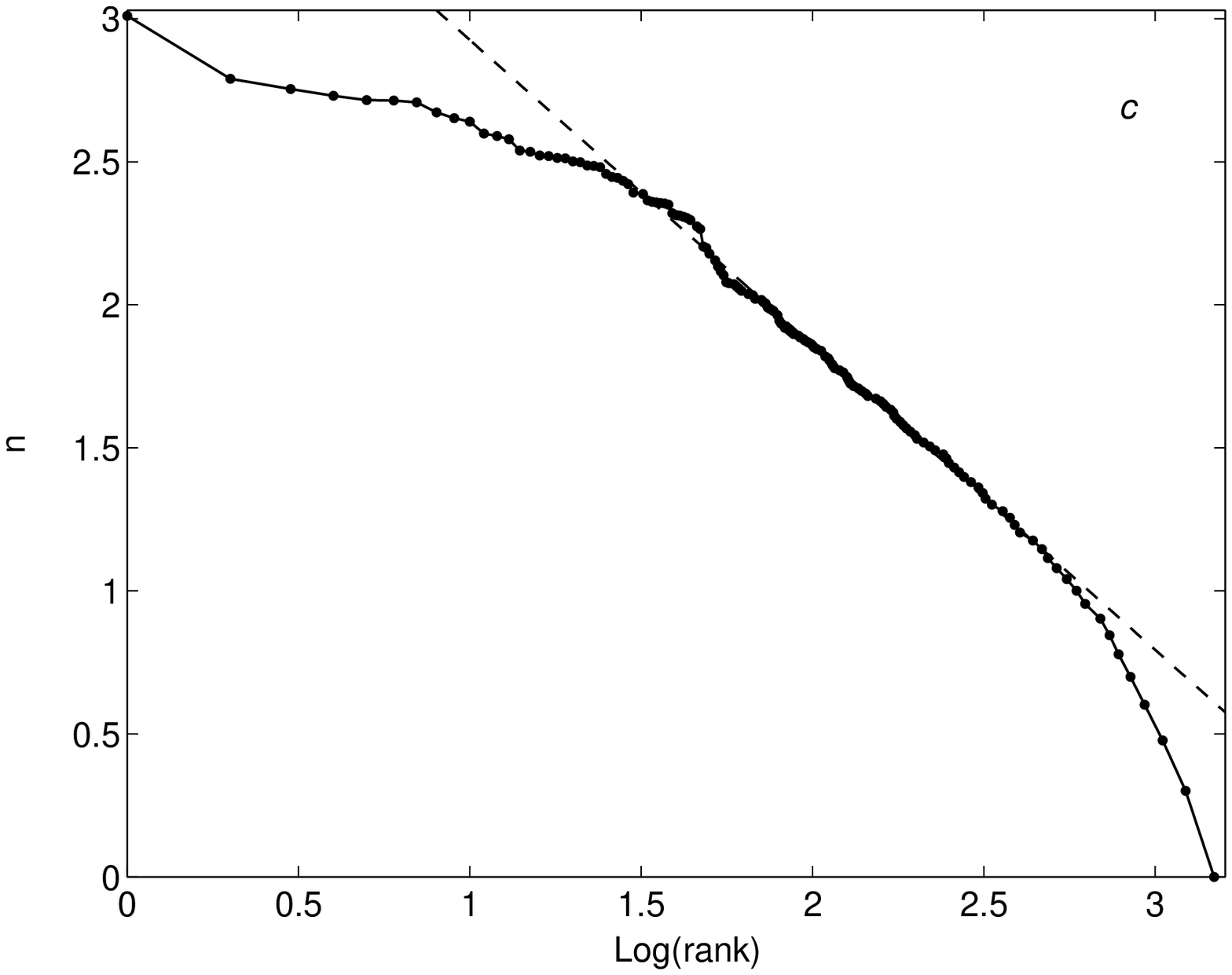}
\includegraphics[height=5cm,width=6.5cm]{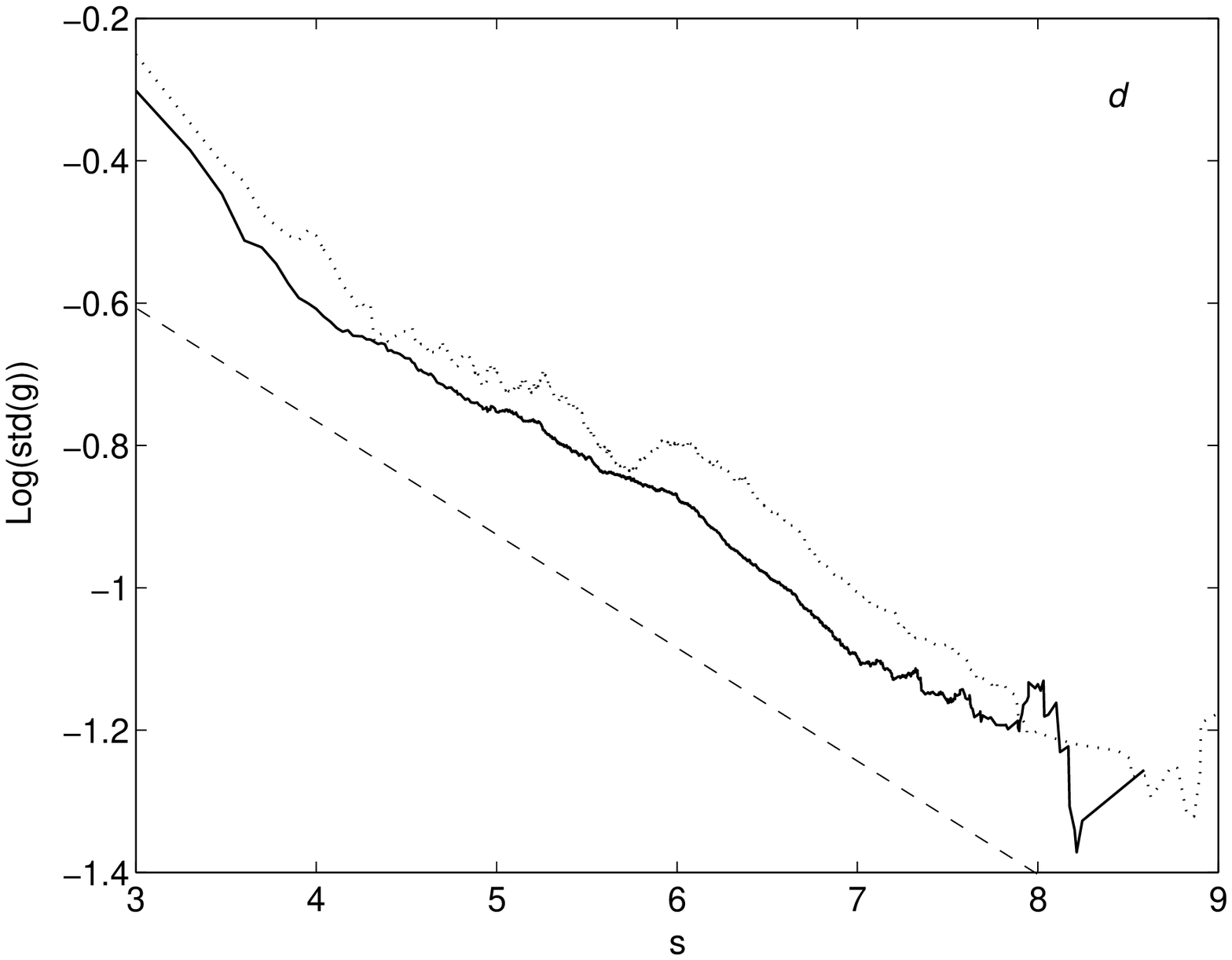}
\end{center}
\caption{Probability distributions and the size-variance
relationships (log10-log10 plots). Fig. 2.a - Size distributions,
products (dots) and firms (circles). Gaussian fit: products\
(.048, 4.34, 1.59); firms (.043, 4.63, 1.97). Fig. 2.b - Product
(dots) and firm (circles) growth distributions. Fig. 2.c - Zipf
plot of the number of products by firm in 2001. Pareto fit: -1.06
($\pm.05$). Fig. 2.d - Size-variance relationship for products
(full line) and firms (dotted line). Linear slope coefficients are
-.167 ($\pm.005)$ for firms
and -.156 ($\pm.012)$ for products ($-.16$ reference line.)}%
\end{figure}

Since a few seminal contributions
\cite{Hymer_Pashigian62,Mansfield62,Simon64}, it is well known
that the standard deviation of firm growth rates tends to decrease
with firm size less rapidly than the square root of size.
Recently, \cite{stanley96} and \cite{sutton02} have provided a
robust characterization to this stylized fact by fitting a power
law relationship of the form $std(g)=cS^{\beta}$ and estimating
the power coefficient $\beta$ in the range $-.15$ to $-.21.$ Fig.
2.d shows similar results for both pharmaceutical products and
companies ($\beta\cong -.16).$ As in \cite{sutton02}, this
departure from the prediction of the Law of Large Numbers cannot
be interpreted as the effect of some form of correlation of growth
rates across constituent units at the level of each firm. In fact,
the mean cross-correlation of products at the firm level is weak
($.07),$ and its effect is too small to account for the flatness
of the scaling relationship. Fig. 2.c. reports the Zipf's plot of
the relationship between $Log(N_{i})$ and the logarithm of the
rank of companies in terms of number of subunits. The distribution
of the number of products can be approximated by a Pareto
distribution with a slope coefficient of $-1.06$ with a departure
in the tails (first 20 companies and small firms with less than 10
products).

As for growth rates, some departures from a pure Gibrat process
can be detected. First, the stationarity of the variance
substantiates a first deviation from a Gibrat growth process,
which predicts a linear increase of the variance with time.
Second, the standard deviation of the growth rates does not
decrease according to $\frac{1}{\sqrt{S}}$. Third, the growth
distributions are non-Gaussian and leptokurtotic.

\section{Scaling properties and proportional growth}

Firms grow through the launch of new products and the growth in
size of existing products. This two-way mechanism  is essential to
characterize the firm growth process, which  is assumed here to be
the outcome of the law of proportional effect applied on sizes and
 number of opportunities. In this section, we investigate how
these two mechanisms affect the shape of the growth distributions
at different levels of aggregation and the scaling relationship.

\begin{figure}[!t]
\begin{center}
\includegraphics[width=6.5 cm]{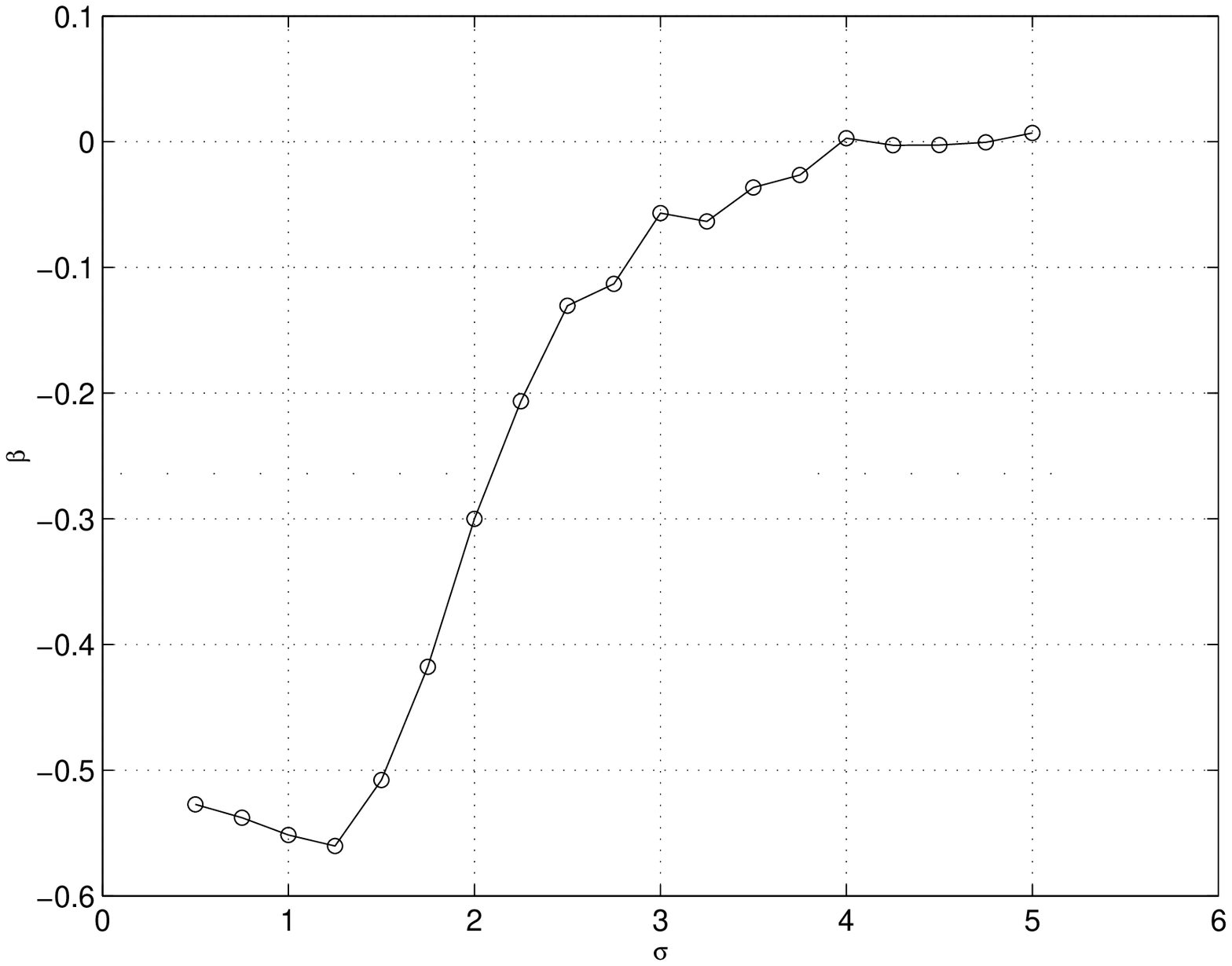}
\includegraphics[width=6.5 cm]{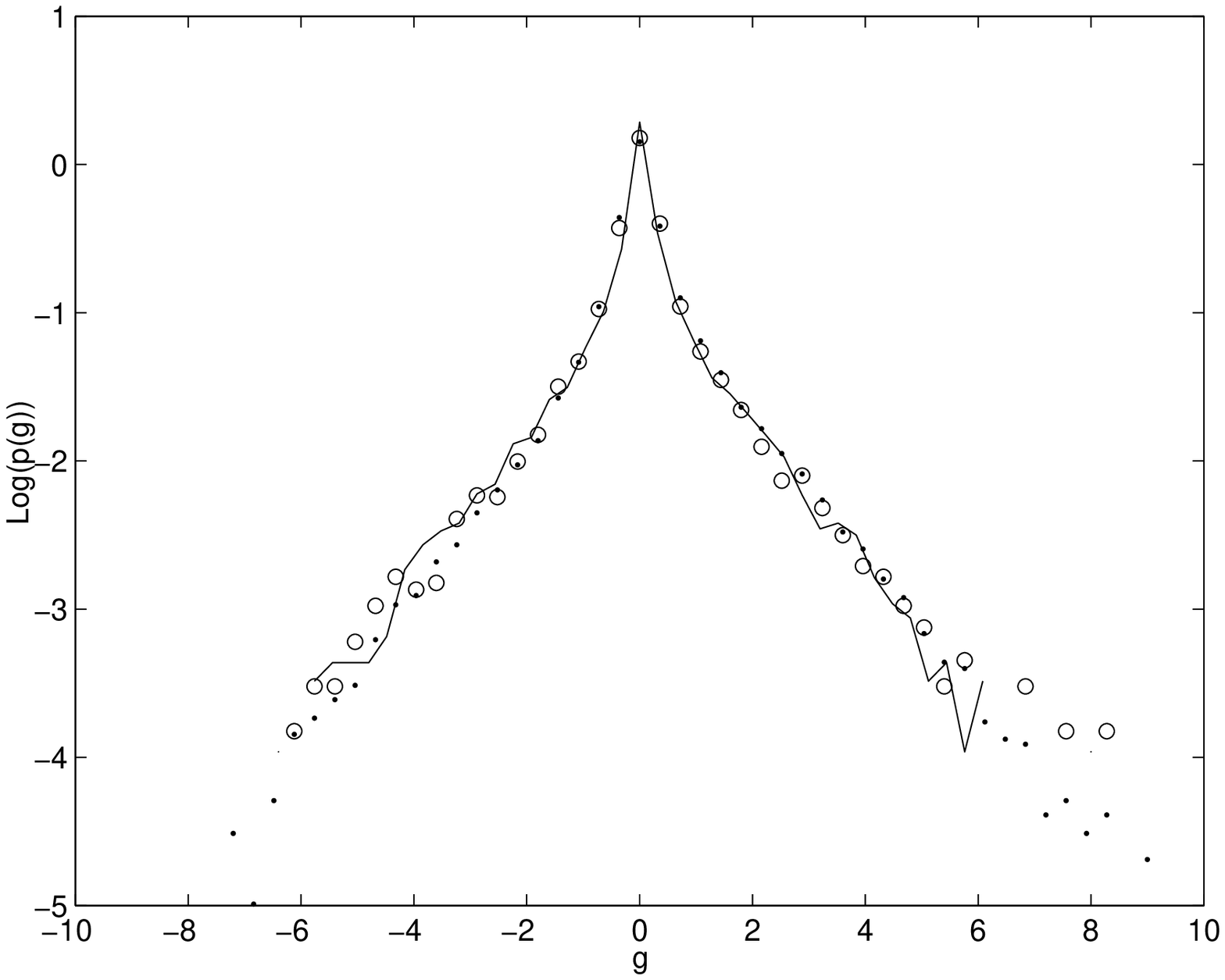}
\end{center}
\caption{Left: Dependence of $\beta$ in the scaling relation $std(g)=S^{\beta
}$ against the coefficient $\sigma$ of the underlying size distribution.
Right: Growth distributions, empirical (dots for products, circles for firms)
and simulative results (solid line).}%
\label{simul}%
\end{figure}

Like in \cite{Ijiri_Simon75}, each distinguishable arrangement of
products at the firm level is assumed to have an equal probability
of occurrence. Assignment or loss of business opportunities to
firms is modelled by randomly selecting firm $i$ proportionally to
the number of its products $N_{i}$. Each time a firm is selected a
new product is given to that firm. In this way, the number of
products assigned represents  the total time of the simulation.
Then, new firms are added at rate $\alpha$, and old products are
removed every $1/\delta$ new product assignments. In line with our
empirical findings, this model leads to a Pareto distribution of
the number of constituent business units \cite{Ijiri_Simon75}.
Once captured, each elementary unit  is assumed to grow in size
following a geometric process which in logs reads
$ds_{t}=adt+bdW_{t}$. We set $a=0$  and $b$ small, sampling the
initial size $S_{0}$ of new products from a sum of lognormal
distributions with mean and standard deviation equal to the
empirically observed values $(3.48$, $2.27)$. The result of the
simulation is plotted in Fig. \ref{simul} together with the
empirical growth distribution for products and firms. It is
obtained by assigning $25,000$ products to $2,000$ firms. Then,
further $10,000$ products are added, with $\alpha=0.01$ and
$\delta=0.05$.

The logarithm of firm size $S_{i}$ at time $t$ is equal to the
logarithm of the sum of the sizes of its constituent components
($s_{i}=\log\left( \sum_{j}S_{j}\right)  $). As we have shown in
Section 2, both product and firm sizes $S$ are approximately
distributed as a log-normal distribution with an upper tail which
decays as a power law. The sum of lognormally distributed random
variables does not have a close form, while several possible
approximations have been proposed for the first two moments, which
involve series evaluations. These estimates are all based on the
approximation that a sum of lognormals is still a lognormal
distribution \cite{Schwartz81}, stable upon aggregation. In fact,
a log-normal distribution $p(S)$ with parameters ($\mu,\sigma$)
behaves as a power law between $S^{-1}$ and $S^{-2}$ for a wide
range of its support $S_{0}<S<S_{0}e^{2\sigma^{2}}$, where $S_{0}$
is a characteristic scale corresponding to the median
\cite{sornette_critical}. Since a decay similar to power-law is
present for a large part of the upper tail, the central limit
theorem does not work effectively. This argument explains the
stability of size distributions upon aggregation. In particular,
simple numerical simulations show that $\log\left(
\sum_{j}S_{j}\right)  $ depicts a Pareto $1/S$ tail, in line with
the empirical distributions.

For a fixed number of products, numerical simulations show that
$p(g)$ is to a good approximation distributed as a Laplace. In
fact, because  logarithms of sum of lognormal distributions in $S$
tend to an exponential in the upper tail for $\log(S)$, the
difference is distributed as a Laplace on the tails. The scaling
relationship is $std(g)=S^{\beta}$, with $\beta$ dependent on the
standard deviation of the underlying size distribution. The
coefficient $\beta$ goes from $-0.5$ for $\sigma$ very small  to
zero for a large $\sigma$. In the parameter range of our empirical
data, $\beta$ is between $-0.1$ and $-0.2$ for a variance of the
underlying size distribution that spans over three orders of
magnitude. This process of size growth by itself has a small
variance and influences the observed growths only in the central
part of the distribution. Moreover, although the variance of $s$
for this process tends to increase linearly in time, we find its
magnitude to be  very small, accounting for the observed
stationarity  of the standard deviation of the log-size $s$ in the
spanned time period. The  shape of the empirical growth
distributions is mostly due to the growth in number of products
and to the distribution of the aggregation process in $N_{i}$
which produces the result plotted in Fig \ref{simul}.

\section{Conclusions}

This paper shows that the framework originally developed by H. A.
Simon and Y. Ijiri can be extended to account for some universal
features in economic and industrial growth, which have been
detected across different domains following \cite{stanley96}. Our
work aims at providing a simplified framework to investigate the
mechanisms that sustain processes of economic growth, in terms of
the static and dynamic relationships between the size of economic
entities and the number and size distribution of their elementary
constituent components. It shows that two multiplicative growth
processes in number of opportunities and size are able to
reproduce most of the salient aspects of the empirical growth
process. In particular, the model predicts that the scaling
relation is stable for a wide range of variances of the underlying
size distribution, and the growth process is stable upon
aggregation. Further research is needed to articulate the
assumptions of the outlined framework in different regimes of
growth, in the direction of building parsimonious and realistic
representations of economic and industrial growth.

\section*{Acknowledgements}

This study was supported by a grant of the Merck Foundation (EPRIS Program).
We thank H.\ E. Stanley, J. Sutton, X. Gabaix, and K. Matia for interesting
discussions.
\bibliographystyle{elsart-num}
\bibliography{../growth}
\end{document}